\documentstyle[11pt,newpasp,twoside,epsf]{article}
\begin{document}
\title{A massive star-forming region in a very early stage of evolution}
\author{K. J. Brooks}
\affil{European Southern Observatory, Casilla 19001, Santiago 19, Chile}
\author{G. Garay, D. Mardones}
\affil{Departamento de Astronom\'{\i}a, Universidad de Chile, Casilla 36-D,
Santiago Chile}
\author{R. P. Norris}
\affil{ Australia Telescope National Facility, P.O. Box 76,
Epping 1710 NSW, Australia}
\author{M. G. Burton}
\affil{School of Physics, University of New South Wales, Sydney 2052, NSW,
Australia}
\begin{abstract}
We present results from a study of two luminous IRAS sources thought
to be young massive star-forming regions and which have no previously
detected radio continuum emission: IRAS 15596-5301 and IRAS 16272-4837. Our
study incorporates sensitive ATCA radio continuum data, SEST 1.2-mm
continuum (using the new SIMBA bolometer) and line data, as well as data
taken from the MSX database. The results show that both sources are
associated with dense molecular cores which appear to host recently formed
massive stars. We argue that IRAS16272 is in a very early stage of
evolution, prior to the formation of an ultra compact HII region and that
IRAS15596 is in a more advanced stage and hosts a cluster of B-type stars.
\end{abstract}

\section{Introduction}

We are undertaking a multi-wavelength study of a sample of 20 luminous IRAS
sources that are thought to be representative of young massive star-forming
regions. The sources in our sample were chosen from the Galaxy-wide survey
of CS(2--1) emission towards 843 IRAS sources with infrared colours typical
of compact HII regions (Bronfman, Nyman, \& May 1996). Each source selected
for our sample showed line profiles indicative of either inward or outward
motions and in some cases broad wings. Their IRAS luminosities were in the
range $2 \times 10^4 - 4 \times 10^5$ L$_{\sun}$, implying that they all
contain at least one embedded massive star ($>$ 8 M$_{\odot}$).

Two sources in our sample, IRAS 15596-5301 and IRAS 16272-4837, were found
to have no detectable radio emission (down to 2 mJy beam$^{-1}$) in the
survey by Walsh et al. (1998), making them candidates for massive stars in
very early stages of evolution when dense material is still falling towards
the protostar and quenching the development of an ultra compact (UC) HII
region (e.g. Yorke 1984). Here we present the findings from our study
towards these two exceptional sources.

Our study involves radio continuum observations using the Australia
Telescope Compact Array (ATCA), 1.2-mm continuum observations using the new
SIMBA 37-channel bolometer array installed at the SEST, as well as a series
of SEST molecular-line observations between 85 and 250 GHz.  We have also
incorporated mid-infrared data obtained with the Midcourse Space Experiment
(MSX) satellite. For IRAS15596 and IRAS16272, the ATCA radio continuum
observations yielded a 1$\sigma$ detection limit at 4.8-GHz of 0.08 mJy
beam$^{-1}$. The synthesized beams of the resultant images (FWHM) were
$\approx$ 2\arcsec. At 1.2-mm (230 GHz) the FWHM beamsize of the SEST is 24\arcsec.

\section{Results \& Discussion}

\begin{figure}
\plotfiddle{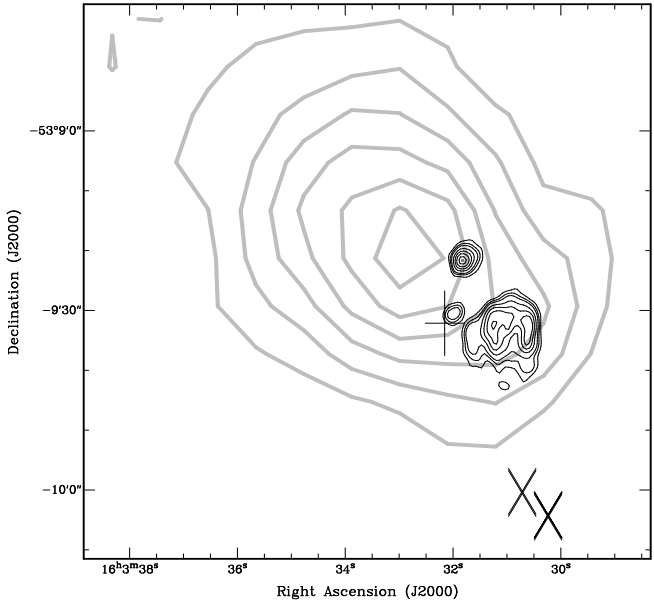}{0.55\textwidth}{0}{55}{55}{-200}{-45}
\caption{{\bf IRAS 15596-5301:} 1.2-mm continuum emission (grey contours)
with levels 1, 2, 3, 4, 5, and 6 times 0.3 Jy beam$^{-1}$. 4.8-GHz
continuum emission (black contours) with levels 6, 9, 12, 18, 24, 30 and 36
times 0.07 mJy beam$^{-1}$. Also shown are the positions of OH masers ($+$)
from Caswell (1998) and 6.7-GHz CH$_3$OH masers ($\times$) from Walsh et
al. (1998, 1997).}
\end{figure}

\subsection{IRAS 15596-5301}

The 1.2-mm continuum emission detected towards IRAS15596 is elongated over
42\arcsec\ (0.94 pc at the distance of 4.6 kpc; Bronfman,
private communication) and has a flux density of 5.8 Jy (see
Fig. 1). Assuming a dust opacity at 1.2 mm of 1 cm$^2$ g$^{-1}$ (Ossenkopf
\& Henning 1994) we derive a mass of $1.4 \times 10^3$ M$_{\sun}$. Results
from the line observations indicate the presence of a molecular gas core
with a rotational temperature of $27 \pm 3$ K (see Fig. 2), a molecular
hydrogen column density N(H$_{2}$) of $8 \times 10^{23}$ cm$^{ -2}$ and
density $n$(H$_2$) of $4 \times 10^5$ cm$^{-3}$.

\begin{figure}
\plottwo{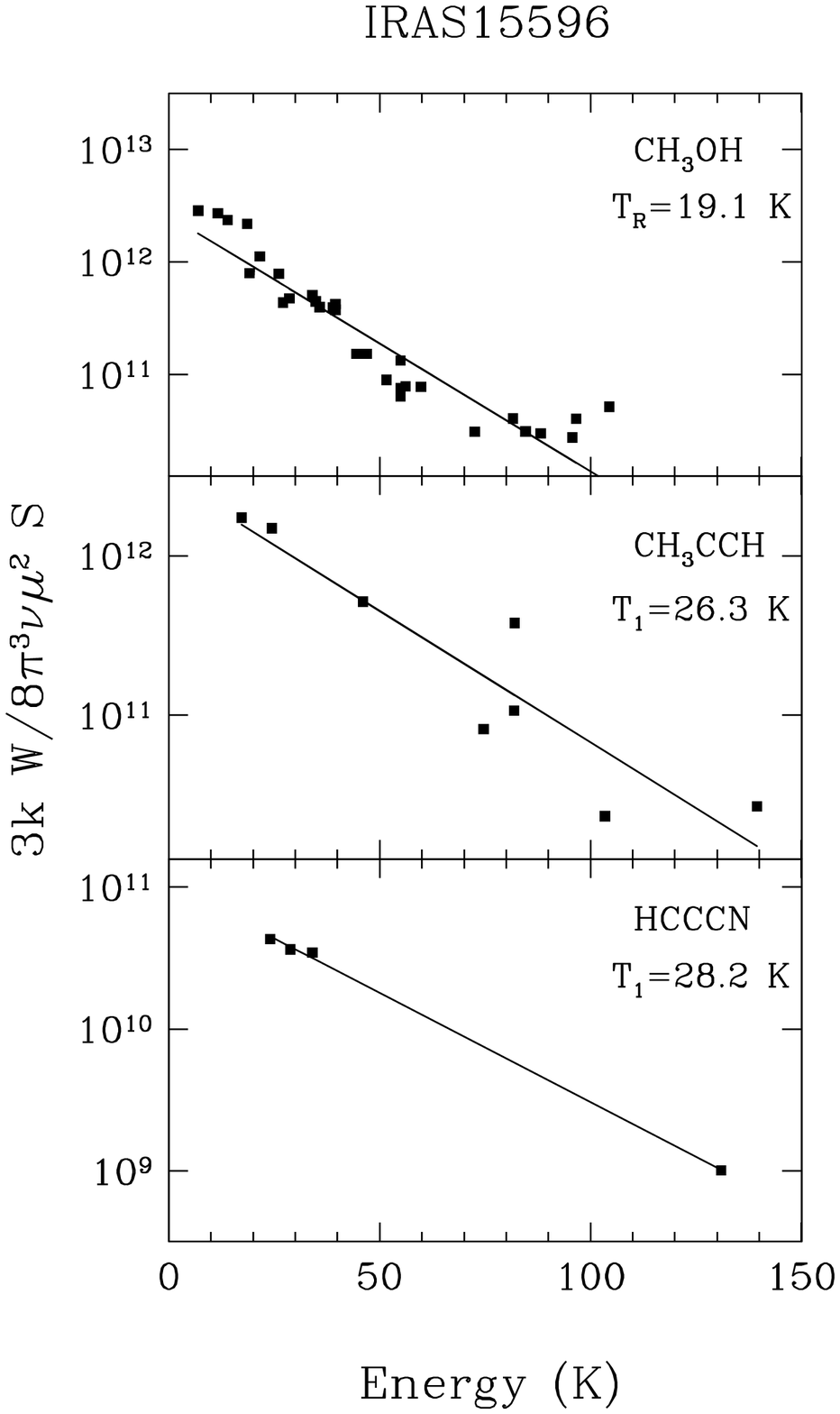}{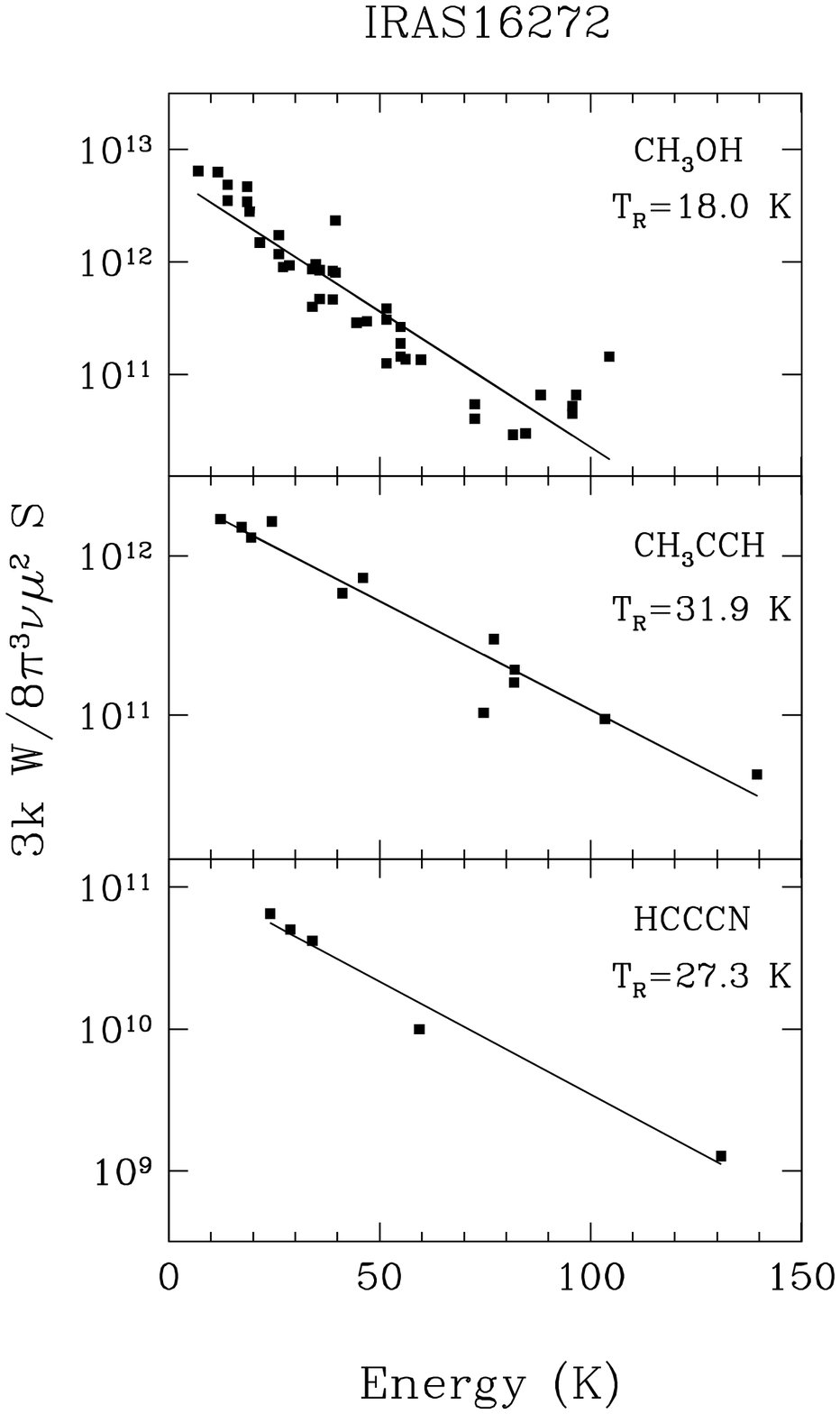}
\caption{Rotational diagrams from selected molecular species observed with
the SEST. The lines correspond to least squares linear fits to the observed
data. The derived values of the rotational temperature (T$_{\rm R}$) are
given in the right corner of each panel.}
\end{figure}

The 4.8-GHz continuum observations reveal three distinct compact sources
with diameters of 0.06 to 0.2 pc. Their peak flux densities are in the
range 1.1 -- 2.8 mJy beam$^{-1}$, close to the sensitivity of the Walsh et
al. (1998) survey. All three sources are located within a region of
30\arcsec\ in diameter and are within the 1.2-mm continuum emission region
(see Fig. 1). Assuming that the three sources are UCHII regions and excited
by individual ZAMS stars, then their flux densities implies the presence of
early B-type stars. The total radio luminosity emitted by this cluster is
$4.1 \times 10^4$ L$_{\sun}$, somewhat lower than the IRAS luminosity of
$6.5 \times 10^4$ L$_{\sun}$. The presence of assorted masers in the
vicinity of IRAS15596 is in accord with this being a massive star forming
region.

From their observed sizes and assuming a sound speed in the ionized gas of
11.4 km s$^{-1}$ we estimate that the three UCHII regions have dynamical
ages between $3 \times 10^3 - 8 \times 10^3$ yr, which might suggest that
they are very young. However, dynamical timescales may not provide
realistic age estimates if the objects have already reached pressure
equilibrium with the surrounding ambient medium. Using the equations
given by Garay \& Lizano (1999) we estimate that a molecular
density of $5 \times 10^5$ cm$^{-3}$ is needed for the three UCHII regions
to be currently pressure confined. This value is similar to that derived
from the molecular line observations. The time needed for the HII regions
to achieve such a pressure equilibrium is between $1 \times 10^5 - 2.5
\times 10^5$ yr, implying that massive star formation started within this
molecular core more than $2.5 \times 10^5$ yr ago.

Although not illustrated here, IRAS15596 is also associated a with compact
mid-infrared source as well as extended MSX A-Band emission ($6.8 - 10.8$
 $\mu$m).

\subsection{IRAS 16272-4837}

\begin{figure}
\plottwo{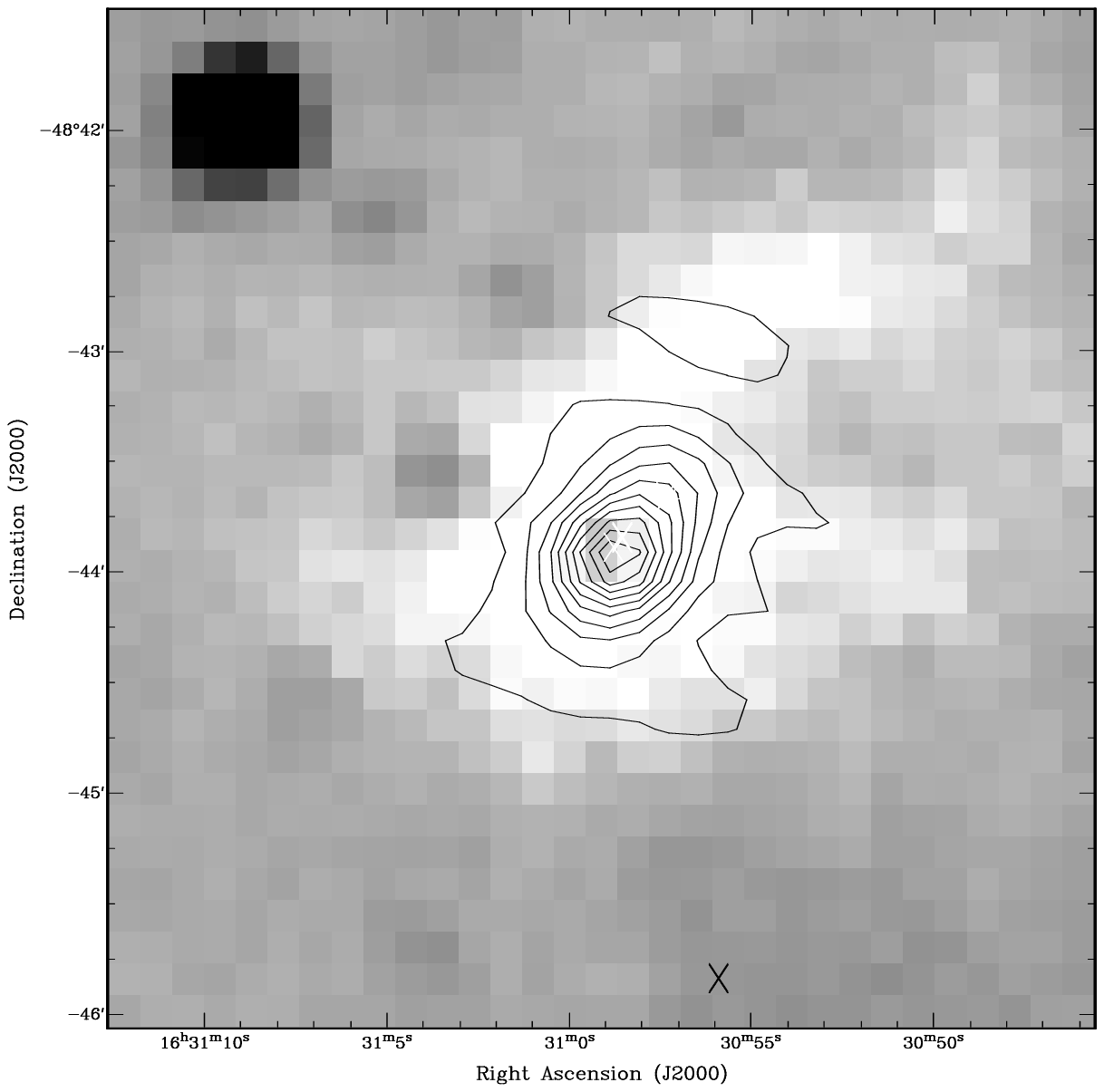}{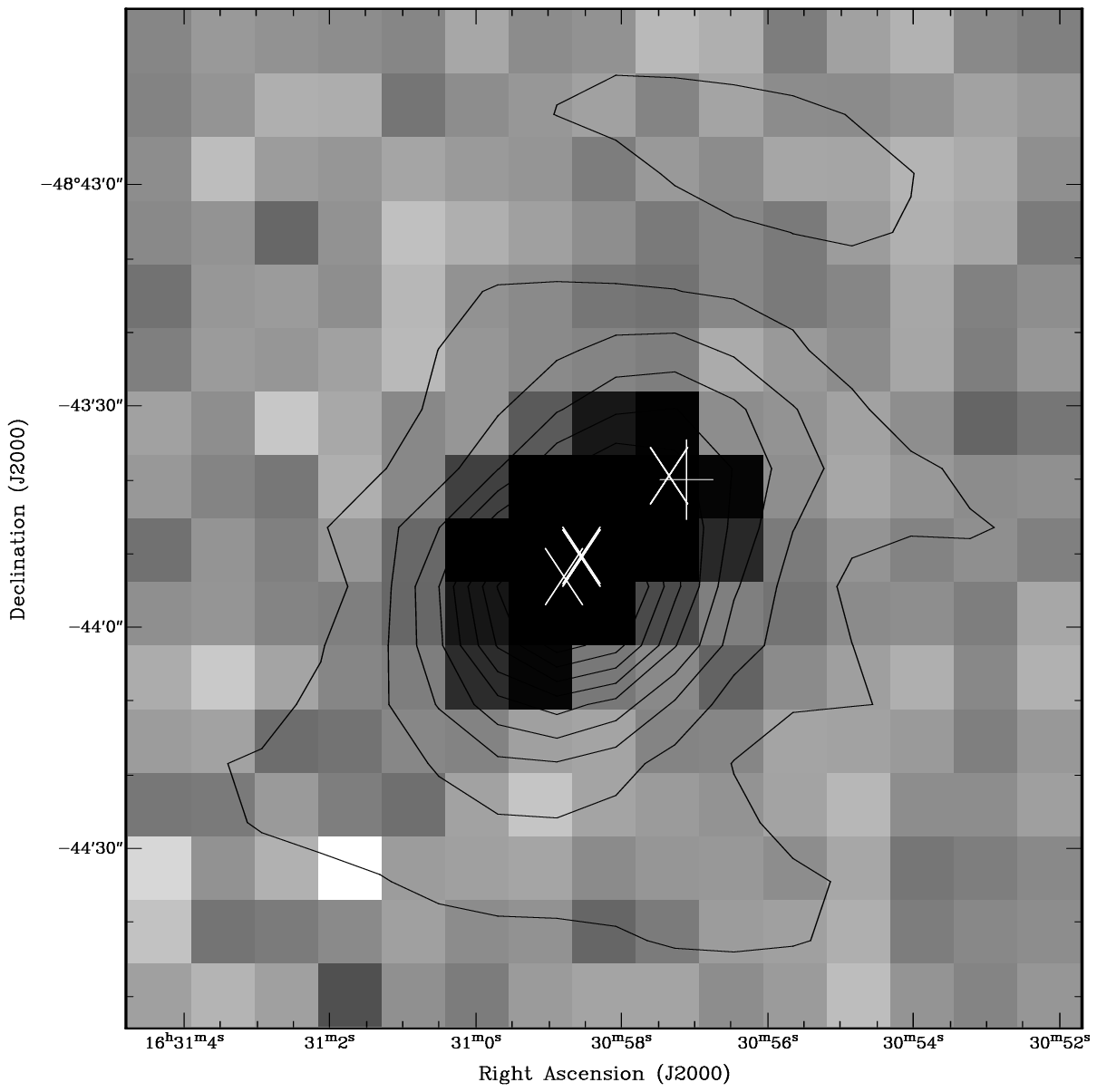}
\caption{{\bf IRAS 16272-4837:} 1.2-mm continuum emission (contours) with
levels 1, 2, 3, 4, 5, 6, 7, 8, 9, 10 times 0.5 Jy beam$^{-1}$. In
reverse grey scale are the MSX images at A-band ({\it left}) and E-band ({\it
right}). Also shown are the maser positions (see caption in Fig. 1).}
\end{figure}

\begin{figure}
\plotfiddle{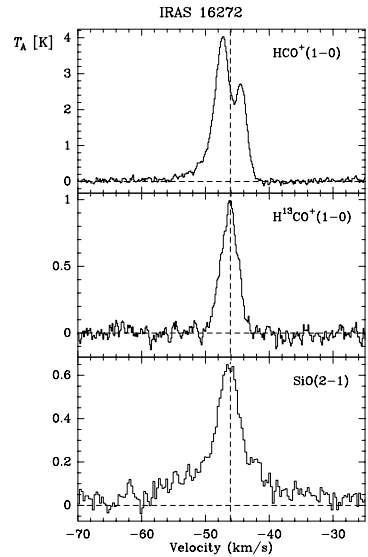}{0.48\textwidth}{0}{32}{32}{-150}{-38}
\caption{A sample of the line profiles observed towards the peak position
of IRAS16272.}
\end{figure}

The 1.2-mm continuum emission detected towards IRAS16272 is elongated over
61\arcsec\ (0.8 pc at the distance of 3.4 kpc; Bronfman private
communication) and has a flux density of 13.8 Jy. Assuming the previously
mentioned dust opacity, we derive a mass of $2 \times 10^3$ M$_{\sun}$. The
line observations indicate the presence of a molecular gas core with a
rotational temperature of 27 $\pm 4$ K (see Fig. 2), N(H$_{2}$) of $2
\times 10^{23}$ cm$^{ -2}$ and $n$(H$_2$) of $2 \times 10^5$ cm$^{-3}$,
similar to IRAS15597. As evident in Fig. 3, the 1.2-mm continuum emission
is associated with a mid-infrared dark cloud seen in absorption in the MSX
A-Band against the emission from the galactic plane.

No 4.8-GHz continuum radio emission was detected to a 3$\sigma$ upper limit
of 0.2 mJy. A non-detection is somewhat surprising given that the IRAS
luminosity of the source is $2.4 \times 10^4$ L$_{\sun}$. If a B0-type star
were responsible for this high luminosity then we expect to measure a radio
flux of $\approx 200$ mJy at optically thin frequencies if embedded in a
uniform density medium. One explanation for the high luminosity and lack of
detected radio emission could be that IRAS16272 is a dense massive
molecular core which hosts a young massive protostar that is still
undergoing an intense accretion phase. Adding support to this hypothesis
are the characteristics of the observed molecular line profiles. For
instance, the behavior of the optically thick HCO$^{+}$(1--0) and optically
thin H$^{13}$CO$^{+}$(1--0) lines shown in Fig. 4 are consistent with
infalling motions. Furthermore, the broad wings evident in the SiO(2--1)
spectrum suggest the presence of outflowing gas, a phenomenon thought to be
closely related to accretion processes.

The notion that IRAS16272 hosts a young massive protostar is also supported
by the presence of assorted maser emission and an emission source seen in
the MSX E-Band ($18.2 - 25.1$ $\mu$m). It is tantalizing to note that this
source appears slightly extended with the maser spots aligned along the
major axis (see Fig. 3).

\end{document}